\documentclass[a4paper,11pt, final]{IEEEtran}
\usepackage{graphics, graphicx}
\usepackage{amssymb,amsmath}
\usepackage{amsmath}
\usepackage[rm]{subfigure}    
\usepackage{threeparttable}   
\usepackage[dvips]{color}
\usepackage{url}   
\usepackage[latin1]{inputenc} 
\usepackage[T1]{fontenc}
\usepackage{algorithm}               
\usepackage{cite}
\usepackage{multirow}
\usepackage[dvips]{color}
\usepackage{lineno}
\usepackage[normalem]{ulem}
\usepackage{soul}  
\usepackage{algorithm}
\usepackage{indentfirst}
\usepackage{setspace}
\usepackage{cases}
\allowdisplaybreaks

\newcommand{\bm}[1]{\mbox{\boldmath{$#1$}}}

\usepackage{tikz}
\definecolor{gold}{rgb}{0.85,.66,0}

\begin{document}
\title{BER Analysis of Multi-Cellular MIMO Systems with Increasing Number of BS Antennas}
\author{José Carlos Marinello, Taufik Abrão\\
\vspace{-4mm}
\thanks{J. C. Marinello is an Master of Science candidate at the Electrical Engineering Department, State University of Londrina,. E-mail: \texttt{\scriptsize zecarlos.ee@gmail.com}}
\thanks{T. Abrão is an Associate Professor at the Electrical Engineering Department, State University of Londrina, PR, Brazil.  E-mail: \texttt{\scriptsize taufik@uel.br}}
}

\maketitle 

\begin{abstract}
In this work, salient characteristics of a wireless communication system deploying a great number of antennas in the base station (BS), namely Massive MIMO system, are investigated. In particular, we found a simple and meaningful relationship that corroborates a fundamental assumption in recent related works: according the number of BS antennas $N$ grows, the product of the small-scale fading channel matrix with its conjugate transpose tends to a scaled identity, with a variability measure inversely proportional to $N$. Furthermore, analysis of the Massive MIMO system downlink is carried out from a bit-error-rate (BER) performance viewpoint, including some realistic adverse effects, such as interference from neighboring cells, channel estimation errors due to background thermal noise, and pilot contamination, which was recently shown to be the only impairment that remains in the MIMO multicell system with infinite number of BS antennas. Our numerical result findings show that, in the same way as with the sum capacity, the pilot contamination also limits the BER performance of a noncooperative multi-cell MIMO system with infinite number of BS antennas.
\end{abstract}


\section{Introduction}
\PARstart{M}{ultiple}-input-multiple-output (MIMO) techniques constitutes one of the key features in most of the recent telecommunications standards, such as WiFi, WiMAX, LTE \cite{hanzo2010_lte_wifi_wimax}, \cite{Li10}, due to the large gains in spectral/energy efficiency they can offer. In particular, multiuser MIMO (MU-MIMO) systems have attracted substantial interest since they can achieve spatial multiplexing gains even when serving single antenna mobile terminals (MT's) \cite{Gesbert07}. Advantages of MU-MIMO also includes a larger robustness to most of propagation limitations present in single-user MIMO, such as antenna correlation or channel rank loss. Even when the channel state information (CSI) of some users are highly correlated, multiuser diversity can be extracted by efficient techniques of scheduling the time/frequency resources, yielding to a better exploitation of the additional degrees of freedom (DoF) propitiated by the antenna array at the BS. From an information theoretic point of view, gains of these systems have been demonstrated in \cite{Tse03}.

In order to fully exploit the advantages of MIMO systems, a certain number of recent technical works have been concerned with the possibility of increasing the number of BS antennas $N$ to infinity. While early papers have focused on the asymptotic limits of such systems for pure academic interest, practical issues of implementation are each time more present on the latest papers, maturing the technology and turning it gradually ready to be considered in next telecommunications standards, such as 5G \cite{Heath14}. It was shown in \cite{Marzetta10} that in a time division duplex (TDD) noncooperative multi-cell MIMO system, that employs uplink training pilots for CSI acquisition, with infinite number of BS antennas, the effects of uncorrelated thermal noise and fast fading are averaged out. The only factor that remains limiting performance in Large MIMO is inter-cell interference, which when associated with the finite time available to send pilot sequences makes the estimated CSI at one BS ``contaminated'' by the CSI of users in adjacent cells, in the so-called pilot contamination effect. This phenomenon  results from unavoidable re-use of reverse-link pilot sequences by terminals in different cells. As a consequence of increasing the number of BS antennas to infinity, the transmit power can be designed arbitrarily small, since interference decreases in the same rate of the desired signal power, i.e., signal-to-interference-plus-noise ratio (SINR) is independent of transmit power \cite{Marzetta10}.

Alternative strategies to achieve better CSI estimates exist, such as a) frequency division duplex (FDD) \cite{Choi13}, in which pilots for CSI acquisition are transmitted in downlink, and fed back to BS in a feedback channel; and  b) network MIMO \cite{Valenzuela06}, where CSI and information data of different coordinated cells are shared among them in a backhaul link, creating a distributed antenna array that serves the users altogether. However, both schemes becomes unfeasible when $N \rightarrow \infty$, since lengths of forward pilot sequences and capacity of backhaul links increase substantially with $N$, respectively \cite{Marzetta10}.

Operating with a large excess of BS antennas compared with the number of terminals $K$ is a challenging but  desirable condition, since some results from random matrix theory become noticeable \cite{Debbah13}, \cite{Matthaiou10}. It is known, for instance, that very tall/wide matrices tend to be very well conditioned, since their singular values distribution appears to be deterministic, showing a stable behavior (low variances) and a relatively  narrow spread \cite{Rusek13}. This phenomenon is quite appreciable to enhance the achievable rates of such systems. Besides, the most simple uplink/downlink techniques, i.e., maximum ratio combining (MRC) and matched filtering (MF) precoding, respectively, becomes optimal \cite{Rusek13}. The savings in energy consumption are also remarkable. In the uplink, it is shown in \cite{Ngo13_EE_SE} that the power radiated by the terminals can be made inversely proportional to $N$, with perfect CSI, or to $\sqrt{N}$, for imperfect CSI, with no reduction in performance, propitiating the implementation of very energy-efficient communication systems.

An interesting investigation about precoding techniques to be deployed in the downlink of the noncooperative multi-cell MU-MIMO systems is performed in \cite{Yang13}. Specifically, authors compared MF precoding, also known as conjugate beamforming, and zero forcing (ZF) beamforming, with respect to net spectral-efficiency and radiated energy-efficiency in a simplified single-cell scenario where propagation is governed by independent Rayleigh fading, and where CSI acquisition and data transmission are both performed during a short coherence interval. It is found that, for high spectral-efficiency and low energy-efficiency, ZF outperforms MF, while at low spectral-efficiency and high energy-efficiency the opposite holds. A similar result is found for the uplink in \cite{Ngo13_EE_SE}, where for low signal-to-noise ratio (SNR), the simple MRC receiver outperforms  the ZF receiver. It can be explained since, for reduced power levels, the cross-talk interference introduced by the inferior maximum-ratio combining receiver eventually falls below the noise enhancement induced by zero forcing and this simple receiver becomes a better alternative. The analog occurs for downlink.

A more rigorous formula for the achievable SINR in Massive MIMO systems is derived in \cite{Fernandes13}; besides, authors discuss an efficient technique for temporally distribute the uplink transmissions of pilot sequences, avoiding simultaneous transmissions from adjacent cells and reducing interference as well, in conjunction with power allocation strategy. The achieved gains in the sum rate are about 18 times, remaining, however, limited to the increasing $N$. On the other hand, a precoding technique that eliminates pilot contamination and leads to unlimited gains with $N \rightarrow \infty$ is proposed in \cite{Ashikhmin12}. However, these gains come at the expense of sharing the information data between base stations, what can overload the backhaul signaling channel for high rate systems, or high number of users per cell.

In \cite{Jose11}, the problem of pilot contamination in noncooperative TDD multi-cell systems is also investigated, and authors derived a closed form expression for a multi-cell minimum mean squared error (MMSE) precoding technique, that depends on the set of training sequences assigned to the users. The technique is obtained as the solution of an objective function consisting of the mean squared error of signals received at the users in the same cell, and the mean squared interference caused at the users in other cells. Hence, achieved gains are resultant of both intra-cell interference and inter-cell interference reduction, while not requiring excessive sharing of information in backhaul signaling channels.

Many practical difficulties arise when trying to design a BS equipped with a massive number of antennas \cite{Rusek13}, such as size limitations, correlation and mutual coupling between antennas\footnote{These effects become more adverse when BS antennas are distributed in more than one dimension \cite{Rusek13}, \cite{Nam13}.}, energy consumption of RF circuits, among others. Due to these limitations, some works have investigated the benefits of a high dimensional MIMO system with a limited number of BS antennas \cite{Hoydis13}, \cite{Caire12}. In \cite{Hoydis13} it was analyzed to which extent the conclusions of Massive MIMO systems hold in a more realistic setting, where $N$ is not extremely large compared to $K$. Authors derived expressions of how many antennas per MT are needed to achieve a given percentage of the ultimate performance limit with infinitely many antennas, and showed by simulations that MMSE/regularized zero forcing (RZF) can achieve the ultimate performance of the simple MRC/MF schemes with a significantly reduced number of antennas. An improved network MIMO architecture is proposed in \cite{Caire12}, that achieves Massive MIMO spectral efficiencies with an order of magnitude fewer antennas. The proposed technique combines cooperation among base stations located in small clusters, ZF multiuser MIMO precoding with suitable inter-cluster interference mitigation constraints, uplink pilot signals allocation and frequency reuse across cells.

When considering that user channels are not independent, but correlated, it can be modelled as a finite-dimensional channel in which the angular domain is separated into a finite number of distinct directions \cite{Ngo13}, or angular bins. This more realistic scenario can be found when BS antennas are not sufficiently well separated, or the propagation environment does not offer rich enough scattering. It is shown in \cite{Ngo13}, for the uplink of a noncooperative multicell MU-MIMO, that the system performance with infinite number of BS antennas but a given finite number $\Omega$ of angular bins is equivalent to the performance of a uncorrelated system with $\Omega$ BS antennas. Indeed, such systems saturates not only due to pilot contamination, but also with the finite dimensionality of the channel. Finally, authors found lower bounds on the capacity of these systems with linear MRC and ZF detectors performed at the receiver side.

In our work, we analytically demonstrate that a fundamental assumption made in most of works cited above holds, i.e., when the number of BS antennas grows, the inner products between propagation vectors for different terminals grow at lesser rates than the inner products of propagation vectors with themselves. In particular, we find that the product of the small-scale fading matrix with its conjugate transpose tends to a scaled identity, with a variance inversely proportional to $N$. Furthermore, as most of previous works have investigated the noncooperative multi-cell TDD MU-MIMO system under the sum-capacity viewpoint, herein we adopt the bit-error-rate performance as salient figure of merit. Our main result indicates that, in the same way as it occurs with the sum-capacity performance, the BER performance of Massive MIMO systems saturates (BER floor effect) with increasing number of BS antennas $N$ when considering pilot contamination.

\section{System Model}\label{sec:model}
In the adopted MIMO system, it is assumed that each one of $L$ base stations with $N$ transmit antennas communicates with $K$ users equipped with a single-antenna MT. Note that the $L$ base stations share the same spectrum and the same set of $K$ pilot signals. We denote the $1 \times N$ channel vector between the $\ell$-th BS and the $k$-th user of $j$-th cell by ${\bf g}_{\ell k j} = \sqrt{\beta_{\ell k j}} {\bf h}_{\ell k j}$, in which $\beta_{\ell k j}$ is the long-term fading power coefficient, that comprises path loss and log-normal shadowing, and ${\bf h}_{\ell k j}$ is the short-term fading channel vector, that follows ${\bf h}_{\ell k j} \sim \mathcal{CN}({\bf 0}, {\bf I}_N)$. Flat fading environment was assumed, where the channel matrix $\bf{H}$ is admitted constant over the entire frame and changes independently from frame to frame (block fading channel assumption). Note that $\beta_{\ell k j}$ is assumed constant for all $N$ BS antennas. Since time division duplex is assumed, reciprocity holds, and thus channel state information is acquired by means of uplink training sequences. During a channel coherence interval, the symbol periods are divided to uplink pilot transmissions, processing, and downlink transmission. Using orthogonal pilot sequences, the number of sequences available is equal to its length, and thus, due to mobility of the users, the number of terminals served by each BS is limited. The same set of $K$ orthogonal pilots of length $K$ is used in all cells; hence, for the $k$-th user of each cell is assigned the sequence ${\bm \psi}_k = [\psi_{1,k} \psi_{2,k} \ldots \psi_{K,k}]$, such that $|\psi_{i,k}| = 1$ and $|{\bm \psi}_{k'}^H {\bm \psi}_k| = \delta_{kk'}$ since the sequences are orthogonal, where $\{\cdot\}^H$ is the conjugate transpose operator, and $\delta_{kk'}$ is the Kronecker's delta, i.e., $\delta_{kk'} = 1$ if $k=k'$ and 0 otherwise.

In the training phase, assuming synchronization in the uplink pilot transmissions, that is the worst case \cite{Marzetta10}, we have that the $N \times K$ received signal at the $\ell$-th BS is:
\begin{equation}\label{eq:rx_pilots}
{\bf Y}_{\ell} = \sum_{j = 1}^{L} {\bf G}_{\ell j}^T \sqrt{{\bm \rho}_j} {\bm \Psi} + {\bf N},
\end{equation}
where ${\bm \rho}_j = {\rm diag}(\rho_{1 j} \, \rho_{2 j} \, \ldots \, \rho_{K j})$, being $\rho_{k j}$ the uplink transmit power of the $k$-th user of $j$-th cell, $\{\cdot\}^T$ is the transpose operator, ${\bf G}_{\ell j} = [{\bf g}_{\ell 1 j}^T \, {\bf g}_{\ell 1 j}^T \, \ldots \, {\bf g}_{\ell K j}^T]^T$, such that ${\bf G}_{\ell j} = {\bm \beta}_{\ell j} {\bf H}_{\ell j}$, ${\bm \beta}_{\ell j} = {\rm diag}(\beta_{\ell 1 j}^{\frac{1}{2}} \, \beta_{\ell 2 j}^{\frac{1}{2}} \, \ldots \, \beta_{\ell K j}^{\frac{1}{2}})$, ${\bf H}_{\ell j} = [{\bf h}_{\ell 1 j}^T \, {\bf h}_{\ell 1 j}^T \, \ldots \, {\bf h}_{\ell K j}^T]^T$, ${\bm \Psi} = [{\bm \psi}_1 \, {\bm \psi}_2 \, \ldots \, {\bm \psi}_K]$, and ${\bf N}$ is a $N \times K$ additive white gaussian noise (AWGN) matrix with zero mean and unitary variance; hence, we can define the uplink SNR as $\text{SNR}^\textsc{ul} = \rho_{k j} \beta_{j k j}$.

To generate the estimated CSI matrix $\widehat{\bf G}_{\ell}$ of their served users, the $\ell$-th BS correlates its received signal matrix with the known pilot sequences:
\begin{equation}\label{eq:rx_pilots_correlator}
\widehat{\bf G}_{\ell}^T = \frac{1}{K} {\bf Y}_{\ell} {\bm \Psi}^H = \sum_{j = 1}^{L} {\bf G}_{\ell j}^T \sqrt{{\bm \rho}_j} + {\bf N}',
\end{equation}
where ${\bf N}'$ is an equivalent AWGN matrix with zero mean and variance $\frac{1}{K}$. Note that the channel estimated by the $\ell$-th BS is contaminated by the channel of users that uses the same pilot sequence in all other cells.

Besides, information transmit symbol vector of the $\ell$-th cell is denoted by ${\bf x}_{\ell} = [x_{1 \ell} \, x_{2 \ell} \, \ldots \, x_{K \ell}]^T$, where $x_{k \ell}$ is the transmit symbol to the $k$-th user of the $\ell$-th cell, and takes a value from the squared quadrature amplitude modulation ($M$-QAM) alphabet; so, the complex-valued symbol (finite) set is given by $\mathcal{S}= \left\{\mathcal{A}+ \sqrt{-1} \cdot \mathcal{A}\right\}$, where the real-valued finite set $\mathcal{A} = \left\{\pm \frac{1}{2}a;\, \pm \frac{3}{2}a;\, \ldots;\, \pm \frac{\sqrt{M}-1}{2}a \right\}$, with $\sqrt{M}$ representing the modulation  order (per dimension) of the corresponding real-valued ASK scheme. The parameter $a =	\sqrt{6/(M-1)}$ is used in order to normalize the power of the complex-valued transmit signals to $1$. Furthermore, we have assumed that the system is determined, i.e., $N\geq K$.
For analysis simplicity, using matrix notation, the $K\times 1$ complex-valued received signal by the users of the $\ell$-th cell is written as:
\begin{equation}\label{eq:rx_C}
{\bf r}_{\ell} = \sum_{j=1}^{L} \sqrt{{\bm \alpha}_j} {\bf G}_{j \ell} {\bf P}_j {\bf x}_j + {\bf n}_{\ell},
\end{equation}
where ${\bm \alpha}_j = {\rm diag}(\alpha_{1 j} \, \alpha_{2 j} \, \ldots \, \alpha_{K j})$, being $\alpha_{k j}$ the downlink transmit power devoted by the $j$-th BS to its $k$-th user, ${\bf P}_j$ denotes the complex valued $N \times K$ precoding matrix of the $j$-th BS, ${\bf n}_{\ell} \sim \mathcal{CN}({\bf 0}, {\bf I}_K)$ represents the AWGN vector with variance $\sigma_n^2 = \frac{1}{2}$ per dimension, which is observed at the $K$ mobile terminals of the ${\ell}$-th cell. Furthermore, the downlink SNR is defined as $\text{SNR}^\textsc{dl} = \frac{\alpha_{k j} \beta_{j k j}}{\log_2 M}$.

\subsection{Matched Filter Beamforming}
The most simple linear precoding technique is the matched filter beamforming, that simply premultiplies the transmit symbol vector by the conjugate transpose of the estimated small-scale fading matrix:
\begin{equation}\label{eq:MF_precoding}
{\bf P}_{\textsc{mf}_{\ell}} = \frac{1}{\sqrt{\gamma_{\ell}}} \widehat{\bf H}_{\ell}^H,
\end{equation}
where $\widehat{\bf H}_{\ell} = {\bm \beta}_{\ell \ell}^{-1} \, \widehat{\bf G}_{\ell} = {\rm diag}(\beta_{\ell 1 \ell}^{\frac{-1}{2}} \, \beta_{\ell 2 \ell}^{\frac{-1}{2}} \, \ldots \, \beta_{\ell K \ell}^{\frac{-1}{2}}) \, \widehat{\bf G}_{\ell}$, and $\gamma_{\ell}$ is a normalization factor to adjust the precoding matrix with the power constraints, such that $\gamma_{\ell} = {\rm{trace}}(\widehat{\bf H}_{\ell}^H \widehat{\bf H}_{\ell})/K$ for this technique.

\subsection{Zero Forcing Beamforming}
The zero forcing beamforming is another prominent linear precoding technique. It is given by the Moore-Penrose pseudo inverse of the estimated small-scale fading channel matrix:
\begin{equation}\label{eq:ZF_precoding}
{\bf P}_{\textsc{zf}_{\ell}} = \frac{1}{\sqrt{\gamma_{\ell}}} \widehat{\bf H}_{\ell}^H \left( \widehat{\bf H}_{\ell} \widehat{\bf H}_{\ell}^H \right)^{-1}.
\end{equation}

As shown in \cite{Rusek13}, in a single-cell MIMO system, the signal received by the $k$-th user can be written as $r_k = \frac{x_k}{\sqrt{\gamma}} + n_k$, evidencing that this technique totally supresses the intra-cell interference at the cost, however, of decreasing the received SNR, since $\gamma_{\ell} = {\rm{trace}}(\widehat{\bf H}_{\ell}^H \widehat{\bf H}_{\ell})^{-1}/K$ for this scheme can assume high values for ill conditioned small-scale fading channel matrices. This effect is quite similar to the noise enhancement of its analogue zero forcing receiver, or decorrelator.

\subsection{Regularized Zero Forcing Beamforming}
Other efficient linear preacoding technique is the regularized zero forcing beamforming \cite{Peel05}:
\begin{equation}\label{eq:RZF_precoding}
{\bf P}_{\textsc{rzf}_{\ell}} = \frac{1}{\sqrt{\gamma_{\ell}}} \widehat{\bf H}_{\ell}^H \left( \widehat{\bf H}_{\ell} \widehat{\bf H}_{\ell}^H  + \zeta {\bf I}_K \right)^{-1},
\end{equation}
in which $\zeta$ is a parameter of the technique that balances interference supression and SNR decrease. Note that RZF can be seen as a generalization of MF and ZF, since RZF reduces itself to these techniques for $\zeta = \infty$ and $\zeta = 0$, respectively. A suitable value for $\zeta$ is suggested in \cite{Rusek13} as $\zeta = \frac{N}{20}$.

\subsection{Minimum Mean Squared Error Beamforming}
The minimum mean squared error beamforming can be seen as a special case of RZF, in which the parameter $\zeta$ is optimized under the objective of minimizing the mean squared error between information transmit symbol vector and the received signal estimated by BS in single-cell systems \cite{Nossek05}. 


As a result, $\zeta_{\textsc{mmse}} = \frac{K \sigma_n^2}{2 \cdot \text{SNR}^\textsc{dl} \cdot \log_2M}$.

\section{Ultimate Limits of Massive MIMO}\label{sec:Massive_MIMO}
Most of the existing theoretical results of Massive MIMO systems are built upon the assumption \cite{Marzetta10}, \cite{Rusek13}, \cite{Fernandes13}:
\begin{equation}
\underset{N \rightarrow \infty}{\rm lim} \frac{1}{N} {\bf H H}^H = {\bf I}_K.
\end{equation}

This assumption is justified since as the number of base station antennas grows, the inner products between propagation vectors for different terminals grow at lesser rates than the inner products of propagation vectors with themselves. In this section, we analitically prove that this assumption is valid, by simply investigating the variance of this product with increasing $N$, and then applying this result to find the asymptotic SINR of the noncooperative multi-cell MIMO system with infinite number of BS antennas.
\subsection{Asymptotic Orthogonality of Small-Scale Fading Matrices}
Defining the matrix ${\bf Q} = \frac{1}{N} {\bf H H}^H$, we have that $q_{i,j} = \frac{1}{N} \sum_{n=1}^{N}h_{i,n} h_{j,n}^*$. In order to determine the variance of the elements in matrix $\bf Q$, the following statistics should be evaluated: $\mathbb{E}[q_{i,i}]$, $\mathbb{E}[q_{i,i}^2]$, $\mathbb{E}[q_{i,j}]$ and $\mathbb{E}[|q_{i,j}|^2]$ for $i\neq j$. Since $|h_{i,j}|$ is a random variable with Rayleigh distribution, we have that $\mathbb{E}[|h_{i,j}|^2] = 2 \sigma^2 = 1$, thus $\sigma = \frac{1}{\sqrt{2}}$, and $\mathbb{E}[|h_{i,j}|^4] = 8 \sigma^4 = 2$.

\begin{enumerate}
\item $\mathbb{E}[q_{i,i}]$
\small
\begin{eqnarray}
&=& \mathbb{E}\left[\frac{1}{N} \sum_{n=1}^{N}h_{i,n} h_{i,n}^*\right] = \mathbb{E}\left[\frac{1}{N} \sum_{n=1}^{N}|h_{i,n}|^2\right] \nonumber\\
&=& \frac{1}{N} \sum_{n=1}^{N}\mathbb{E}\left[|h_{i,n}|^2\right] = \frac{1}{N} \sum_{n=1}^{N}1 = 1. \nonumber
\end{eqnarray}
\normalsize
\item $\mathbb{E}[q_{i,i}^2]$
\scriptsize
\begin{align}
&= \mathbb{E}\left[\frac{1}{N^2} \sum_{n_1=1}^{N}h_{i,n_1} h_{i,n_1}^* \sum_{n_2=1}^{N}h_{i,n_2} h_{i,n_2}^*\right] \nonumber\\
&= \frac{1}{N^2} \sum_{n_1=1}^{N}\mathbb{E}\left[ |h_{i,n_1}|^2 \sum_{n_2=1}^{N} |h_{i,n_2}|^2\right] \nonumber\\
&= \frac{1}{N^2} \sum_{n_1=1}^{N}\mathbb{E}\left[ |h_{i,n_1}|^4 \right] + \frac{1}{N^2} \sum_{n_1=1}^{N}\mathbb{E}\left[ |h_{i,n_1}|^2 \sum_{\substack{n_2 = 1 \\ n_2 \neq n_1}}^{N} |h_{i,n_2}|^2\right] \nonumber\\
&= \frac{1}{N^2} \sum_{n_1=1}^{N}\mathbb{E}\left[ |h_{i,n_1}|^4 \right] \nonumber\\
&\phantom{{}=1} + \frac{1}{N^2} \sum_{n_1=1}^{N}\mathbb{E}\left[ |h_{i,n_1}|^2 \right] \sum_{\substack{n_2 = 1 \\ n_2 \neq n_1}}^{N} \mathbb{E}\left[|h_{i,n_2}|^2\right] \nonumber\\
&= \frac{1}{N^2} \sum_{n_1=1}^{N} 2 + \frac{1}{N^2} \sum_{n_1=1}^{N} 1 \sum_{\substack{n_2 = 1 \\ n_2 \neq n_1}}^{N} 1 \nonumber\\
&= \frac{1}{N^2} 2N + \frac{1}{N^2} N (N-1) = \frac{N+1}{N} \nonumber
\end{align}
\normalsize
\item $\underset{i\neq j}{\mathbb{E}[q_{i,j}]}$ 
\small
\begin{align}
&= \mathbb{E}\left[\frac{1}{N} \sum_{n=1}^{N}h_{i,n} h_{j,n}^*\right] = \frac{1}{N} \sum_{n=1}^{N}\mathbb{E}\left[h_{i,n}\right]\mathbb{E}\left[h_{j,n}^*\right] = 0. \nonumber
\end{align}
\normalsize
\item $\underset{i\neq j}{\mathbb{E}[|q_{i,j}|^2]}$ 
\scriptsize
\begin{align}
&= \mathbb{E}\left[q_{i,j} q_{i,j}^*\right] = \frac{1}{N^2} \mathbb{E}\left[\sum_{n_1=1}^{N}h_{i,n_1} h_{j,n_1}^* \sum_{n_2=1}^{N}h_{i,n_2}^* h_{j,n_2} \right] \nonumber\\
&= \frac{1}{N^2} \sum_{n_1=1}^{N} \mathbb{E}\left[|h_{i,n_1}|^2 |h_{j,n_1}|^2 + h_{i,n_1} h_{j,n_1}^* \sum_{\substack{n_2 = 1 \\ n_2 \neq n_1}}^{N}h_{i,n_2}^* h_{j,n_2} \right] \nonumber\\ 
&= \frac{1}{N^2} \sum_{n_1=1}^{N} \mathbb{E}\left[|h_{i,n_1}|^2\right] \mathbb{E}\left[|h_{j,n_1}|^2\right] \nonumber\\
&\phantom{{}=1} + \frac{1}{N^2} \sum_{n_1=1}^{N} \mathbb{E}\left[h_{i,n_1}\right] \mathbb{E}\left[h_{j,n_1}^*\right] \sum_{\substack{n_2 = 1 \\ n_2 \neq n_1}}^{N}\mathbb{E}\left[h_{i,n_2}^* \right] \mathbb{E}\left[ h_{j,n_2} \right] \nonumber\\ 
&= \frac{1}{N^2} \sum_{n_1=1}^{N}1 = \frac{1}{N}. \nonumber
\end{align}
\normalsize
\end{enumerate}
Since $q_{i,i}$ is a real random variable, we have that ${\rm var}(q_{i,i}) = \mathbb{E}[q_{i,i}^2] - \mathbb{E}[q_{i,i}]^2 = \frac{1}{N}$, where ${\rm var}(\cdot)$ is the variance operator. On the other hand, as $q_{i,j}$, for $i \neq j$, is a complex random variable, we have that ${\rm var}(q_{i,j}) = \mathbb{E}\left[(q_{i,j} - \mathbb{E}[q_{i,j}])(q_{i,j} - \mathbb{E}[q_{i,j}])^*\right] = \mathbb{E}[\left|q_{i,j} - \mathbb{E}[q_{i,j}]\right|^2] = \mathbb{E}[|q_{i,j}|^2] = \frac{1}{N}$.

\subsection{Ultimate SINR With Infinite Number of BS Antennas}
For the MF beamforming, from \eqref{eq:rx_pilots_correlator}, \eqref{eq:rx_C}, and \eqref{eq:MF_precoding}, we have that the vector of signal received by users (downlink) at the $\ell$-th cell ($\ell=1,\ldots,L$) is:
\footnotesize
\begin{equation}\label{eq:rx_MF}
{\bf r}_{\ell} = \sum_{j=1}^{L} \frac{1}{\sqrt{\gamma_j}} \sqrt{{\bm \alpha}_j} {\bf G}_{j \ell} {\bm \beta}_{\ell \ell}^{-1} \left(\sum_{i = 1}^{L} {\bf G}_{\ell i}^H \sqrt{{\bm \rho}_i} + {\bf N}'^*\right) {\bf x}_j + {\bf n}_{\ell}.
\end{equation}
\normalsize

It is demonstrated in \cite{Fernandes13} that the downlink SINR of the $k$-th user of the $\ell$-th cell in this system converges in the limit of infinite $N$ to:
\begin{equation}\label{eq:ultimate_SINR}
\text{SINR}^\textsc{dl}_{k \ell} = \frac{\alpha_{k \ell} \beta^2_{\ell k \ell}/\nu^2_{k \ell}}{\sum_{\substack{j = 1 \\ j \neq \ell}}^L  \alpha_{k j} \beta^2_{j k \ell}/\nu^2_{k j}},
\end{equation}
where the asymptotic power normalization factors $\nu_{k j} = \sum_{i=1}^{L} \rho_{k i} \beta_{j k i} + \frac{1}{K}$. Note that this limit depends mainly on the large scale fading coefficients $\beta_{j k i}$, which are related to the spatial distribution of the users on the different cells. In addition, if one assumes power uniformly distributed between users in the same cell and the same power normalization factors $\nu_{k j}$ as in \cite{Marzetta10}, equation \eqref{eq:ultimate_SINR} simplifies to:
\begin{equation}\label{eq:ultimate_SINR_simp}
\text{SINR}^\textsc{dl}_{k \ell} = \frac{\beta^2_{\ell k \ell}}{\sum_{\substack{j = 1 \\ j \neq \ell}}^L \beta^2_{j k \ell}}.
\end{equation}

\section{Numerical Results}\label{sec:results}

\begin{figure}[!htbp]
\centering
\includegraphics[width=.49\textwidth]{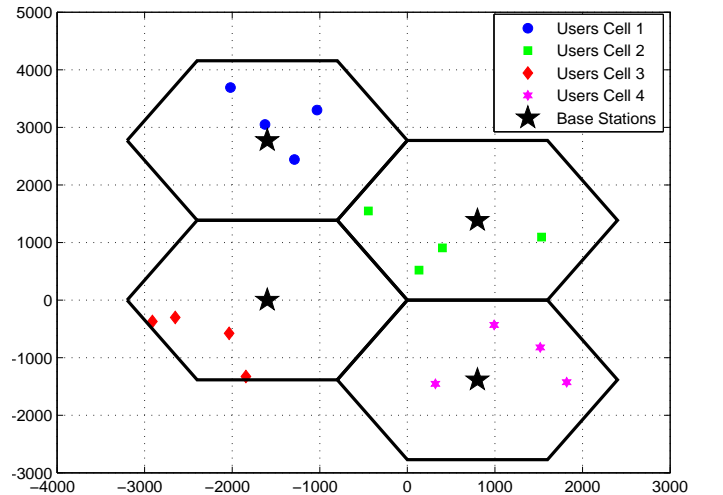}
\vspace{-8mm}
\caption{A random realization for the spatial distribution of mobile users in a multi-cell MIMO system scenario with $L = 4$ and $K=4$.}
\label{fig:spatial_distribution}
\end{figure}

We adopt in our simulations a multi-cell scenario with $L = 4$ hexagonal cells, with radius 1600m, where $K=4$ users are uniformly distributed in its interior, except in a circle of 100m radius around the cell centered BS. A single realization of our spatial model is shown in Fig. \ref{fig:spatial_distribution}.

Besides, we model the log normal shadowing with a standard deviation of 8dB, and the path loss decay exponent equal to 3.8. We simulated two main scenarios of interest from the BER performance metric: in the first, the number of users grows at the same rate as the number of BS antennas, i.e., $N = K$ holds; while in the second, $K$ remains fixed while $N$ grows to infinity. Furthermore, in the simulations setup, we have considered unitary frequency-reuse factor, 4-QAM modulation and $\text{SNR}^\textsc{dl} = 10$dB. It is important to note that results in this section were obtained via Monte-Carlo simulation method.

\subsection{BER Performance under Perfect CSI}
\begin{figure}[!htbp]
\centering
\includegraphics[width=.49\textwidth]{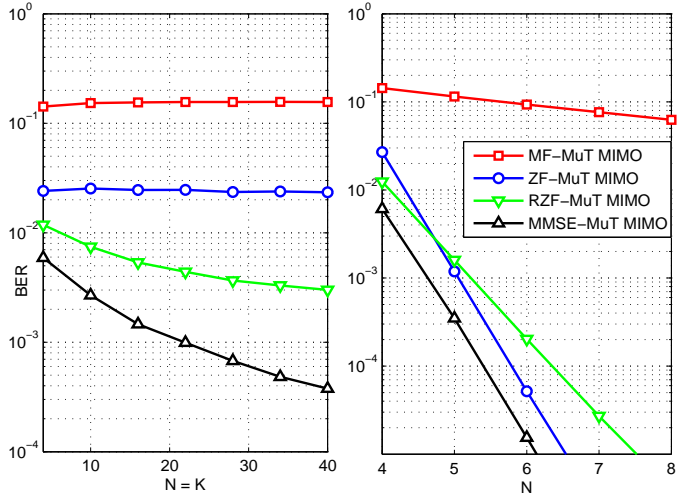}
\vspace{-4mm}
\caption{BER performance of the investigated precoding techniques, for 4-QAM and $\text{SNR}^\textsc{dl} = 10$ dB, with the increasing system dimension: a) $N = K$; b) $K = 4$. Single-cell analysis with perfect CSI.}
\label{fig:ber_single_cell}
\end{figure}

Fig. \ref{fig:ber_single_cell} compares the BER performance for both scenarios in a single-cell systems with perfect CSI. It can be seen for the first scenario ($N=K$) that performance of MF and ZF transmitters stay unchanged with the system dimension increasing, while RZF and MMSE have their performance improved. 
On the other hand, in the second scenario ($K$ fixed and $N$ increasing), all the techniques' performances are improved, and one can see that MMSE and ZF fastly achieves very small error rates, while for RZF and MF, in this order, this occurs more slowly, meaning that the achieved diversity gain for the MF-MuT MIMO system is quite modest.

\begin{figure}[!htbp]
\centering
\includegraphics[width=.49\textwidth]{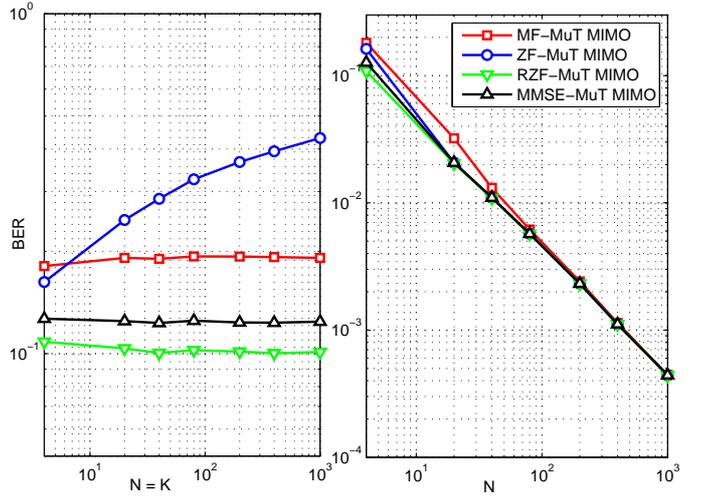}
\vspace{-4mm}
\caption{BER performance of the investigated precoding techniques, for 4-QAM and $\text{SNR}^\textsc{dl} = 10$ dB, with the increasing system dimension: a) $N = K$; b) $K = 4$. Multi-cell analysis with perfect CSI.}
\label{fig:ber_multi_cell_pcsi}
\end{figure}

The same comparison is shown in Fig. \ref{fig:ber_multi_cell_pcsi}, but now for a multi-cell configuration, where inter-cell interference accounts, with perfect CSI. One can see that the considered Massive MIMO precoding techniques result in high bit error rates for the first scenario, that remain fixed with the system dimension increasing, except for ZF, that has its performance getting even worse for $N=K$ growing substantially. For the second scenario, all the techniques improve their performances with increasing $N$; an interesting result is that these performances nearly igualate themselves for $N > 100$, and that RZF surpasses MMSE in terms of BER performance, since the last technique does not account inter-cell interference in its derivation.

\subsection{BER Performance under CSI Estimation Errors}
\begin{figure}[!htbp]
\centering
\includegraphics[width=.49\textwidth]{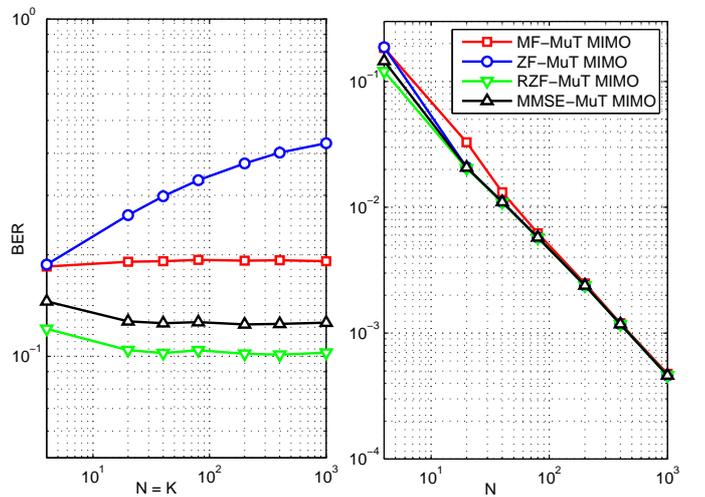}
\vspace{-4mm}
\caption{BER performance of the investigated precoding techniques, for 4-QAM and $\text{SNR}^\textsc{dl}= 10$ dB, with the increasing system dimension: a) $N = K$; b) $K = 4$. Multi-cell analysis with $\text{SNR}^\textsc{ul} = 10$ dB.}
\label{fig:ber_multi_cell_csi_n}
\end{figure}

When considering the multicell scenario but now with imperfect CSI, that is obtained by means of uplink training sequences transmitted with $\text{SNR}^\textsc{ul} = 10$dB, but neglecting the pilot contamination effect, Fig. \ref{fig:ber_multi_cell_csi_n} shows the achieved BER performances. It can be seen that the behavior of the techniques for both scenarios is almost the same than for perfect CSI, as shown in Fig. \ref{fig:ber_multi_cell_pcsi}. Therefore, we can conclude that, when dealing with inter-cell interference, effect of uncorrelated noise in the estimated channel little degrades the system performance.

\begin{figure}[!htbp]
\centering
\includegraphics[width=.49\textwidth]{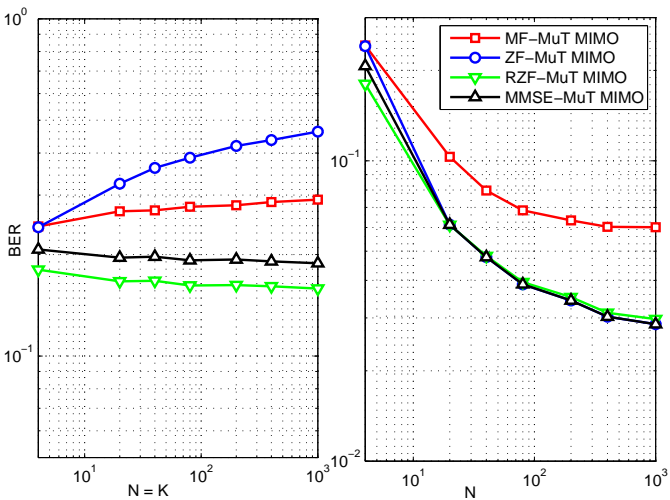}
\vspace{-4mm}
\caption{BER performance of the investigated precoding techniques, for 4-QAM and $\text{SNR}^\textsc{dl} = 10$ dB, with the increasing system dimension: a) $N = K$; b) $K = 4$. Multi-cell analysis with $\text{SNR}^\textsc{ul} = 10$ dB and pilot contamination.}
\label{fig:ber_multi_cell_csi_npc}
\end{figure}

The main result of this paper can be seen when comparing Fig. \ref{fig:ber_multi_cell_csi_n} with Fig. \ref{fig:ber_multi_cell_csi_npc}, that shows the performance of the two scenarios in the complete system model, including pilot contamination. While for the first scenario the same behavior of the BER performances is found, but with even higher error levels, one can see for the second scenario that, when taking pilot contamination into account, the BER performance of the considered precoding techniques saturates with $N \rightarrow \infty$ (BER floor effect), in contrast to Fig. \ref{fig:ber_multi_cell_csi_n}.b. Indeed, one could already expect this result, since errors are induced by noise, intra-cell interference, as well as by inter-cell interference, which can be compared with the desired signal power by means of the SINR. It was already shown in previous works \cite{Marzetta10} and \cite{Fernandes13} that noise and intra-cell interference are averaged out in the MIMO system with very large number of BS antennas, being the inter-cell interference the only remaining impairment, due to pilot contamination. This phenomenom leads to the saturation of the SINR, and, as shown herein, BER performance as well. Besides, the different BER levels between ZF, RZF, MMSE compared to MF can be justified, as explained in \cite{Rusek13} and \cite{Hoydis13}, since the first techniques converge to its asymptotic limits with fewer BS antennas than the last one, although these limits are usually close.

\section{Conclusion}\label{sec:concl}
In this work we have demonstrated that the breakthrough results of Massive MIMO systems downlink can also be found from the bit-error-rate performance perspective. Two scenarios of great interest have been numerically analysed: when the number of users grows in the same rate as the number of BS antennas, and when the number of BS antennas grows keeping fixed the number of users covered in each cell. We found that the first scenario is unable to operate with acceptable performance in multi-cellular systems when considering the linear precoding techniques investigated herein. On the other hand, it was found that the performance of the linear precoding techniques evaluated in the second scenario could be made as good as is wanted, by scaling up the number of BS antennas, if the effect of pilot contamination has not been taken into account. This effect saturates the achievable SINR in the asymptotic limit of BS antennas $N\rightarrow \infty$, as shown in previous works, and consequently the BER performance becomes saturated as well. Furthermore, we proved a widely used assumption that the product of the small-scale fading channel matrix with its conjugate transpose tends to a scaled identity matrix as $N \rightarrow \infty$, and found a closed form expression for the variance of this product. This last result is quite useful, since one can determine with which reliability the assumption is true according the number of antennas deployed in the BS.

 \section*{Acknowledgement}
 This work was supported in part by the National Council for Scientific and Technological Development (CNPq) of Brazil under Grants 202340/2011-2, 303426/2009-8 and in part by Londrina State University - Paraná State Government (UEL).


\begin{IEEEbiography}[{\includegraphics[width=1in,height=1.25in,clip,keepaspectratio]{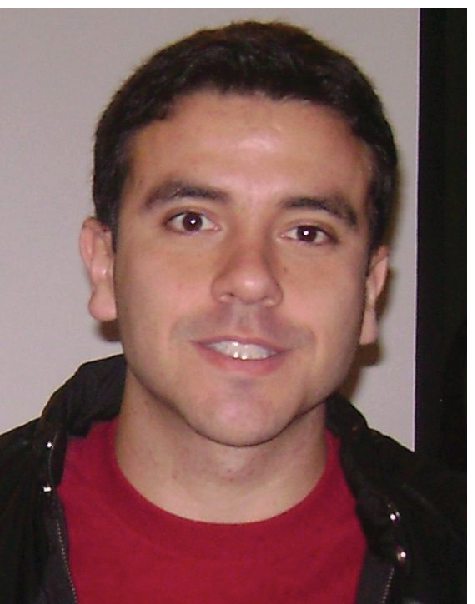}}]
{José Carlos Marinello} received his B.S. in Electrical Engineering (Summa Cum Laude) from Londrina State University, PR, Brazil, in December 2012. He is currently working towards his M.S. and Ph.D. at Londrina State University, Paraná, Brazil. His research interests include physical layer aspects, specially heuristic and convex optimization of 3G and 4G MIMO systems.
\end{IEEEbiography}

\begin{IEEEbiography}[{\includegraphics[width=1in,height=1.25in,clip,keepaspectratio]{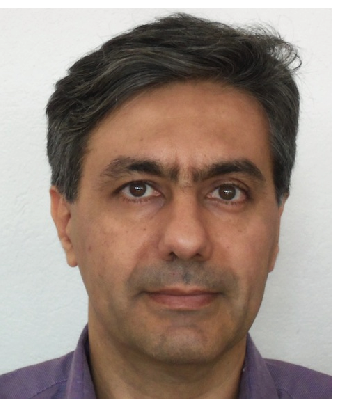}}]
{Taufik Abrão}  (SM'12) received the B.S., M.Sc., and Ph.D. degrees in electrical engineering from the Polytechnic School of the University of São Paulo, São Paulo, Brazil, in 1992, 1996, and 2001, respectively. Since March 1997, he has been with the Communications Group, Department of Electrical Engineering, Londrina State University, Londrina, Brazil, where he is currently an Associate Professor of Communications engineering. In 2012, he was an Academic Visitor with the Communications, Signal Processing and Control Research Group, University of Southampton, Southampton, U.K. From 2007 to 2008, he was a Post-doctoral Researcher with the Department of Signal Theory and Communications, Polytechnic University of Catalonia (TSC/UPC), Barcelona, Spain. He has participated in several projects funded by government agencies and industrial companies. He is involved in editorial board activities of six journals in the communication area and he has served as TCP member in several symposium and conferences. He has been served as an Editor for the IEEE COMMUNICATIONS SURVEYS \& TUTORIALS since 2013. He is a member of SBrT and a senior member of IEEE. His current research interests include communications and signal processing, specially the multi-user detection and estimation, MC-CDMA and MIMO systems, cooperative communication and relaying, resource allocation, as well as heuristic and convex optimization aspects of 3G and 4G wireless systems. He has co-authored of more than 170 research papers published in specialized/international journals and conferences.
\end{IEEEbiography}


\begin{thebibliography}{10}

\bibitem{hanzo2010_lte_wifi_wimax}
L.~Hanzo, Y.~Akhtman, L.~Wang, and M.~Jiang, \emph{MIMO-OFDM for LTE, WiFi and
  WiMAX: Coherent versus Non-coherent and Cooperative Turbo
  Transceivers}.\hskip 1em plus 0.5em minus 0.4em\relax IEEE Press and John
  Wiley \& Sons, November 2010.

\bibitem{Li10}
Q.~Li, G.~Li, W.~Lee, M.~il~Lee, D.~Mazzarese, B.~Clerckx, and Z.~Li, ``Mimo
  techniques in wimax and lte: a feature overview,'' \emph{Communications
  Magazine, IEEE}, vol.~48, no.~5, pp. 86--92, May 2010.

\bibitem{Gesbert07}
D.~Gesbert, M.~Kountouris, R.~Heath, C.-B. Chae, and T.~Salzer, ``Shifting the
  mimo paradigm,'' \emph{Signal Processing Magazine, IEEE}, vol.~24, no.~5, pp.
  36--46, Sept 2007.

\bibitem{Tse03}
P.~Viswanath and D.~Tse, ``Sum capacity of the vector gaussian broadcast
  channel and uplink-downlink duality,'' \emph{Information Theory, IEEE
  Transactions on}, vol.~49, no.~8, pp. 1912--1921, Aug 2003.

\bibitem{Heath14}
F.~Boccardi, R.~Heath, A.~Lozano, T.~Marzetta, and P.~Popovski, ``Five
  disruptive technology directions for 5g,'' \emph{Communications Magazine,
  IEEE}, vol.~52, no.~2, pp. 74--80, February 2014.

\bibitem{Marzetta10}
T.~Marzetta, ``Noncooperative cellular wireless with unlimited numbers of base
  station antennas,'' \emph{IEEE Transactions on Wireless Communications},
  vol.~9, no.~11, pp. 3590--3600, 2010.

\bibitem{Choi13}
J.~Choi, Z.~Chance, D.~Love, and U.~Madhow, ``Noncoherent trellis coded
  quantization: A practical limited feedback technique for massive mimo
  systems,'' \emph{Communications, IEEE Transactions on}, vol.~61, no.~12, pp.
  5016--5029, December 2013.

\bibitem{Valenzuela06}
M.~Karakayali, G.~Foschini, and R.~Valenzuela, ``Network coordination for
  spectrally efficient communications in cellular systems,'' \emph{Wireless
  Communications, IEEE}, vol.~13, no.~4, pp. 56--61, Aug 2006.

\bibitem{Debbah13}
R.~Couillet and M.~Debbah, ``Signal processing in large systems: A new
  paradigm,'' \emph{Signal Processing Magazine, IEEE}, vol.~30, no.~1, pp.
  24--39, Jan 2013.

\bibitem{Matthaiou10}
M.~Matthaiou, M.~McKay, P.~Smith, and J.~Nossek, ``On the condition number
  distribution of complex wishart matrices,'' \emph{Communications, IEEE
  Transactions on}, vol.~58, no.~6, pp. 1705--1717, June 2010.

\bibitem{Rusek13}
F.~Rusek, D.~Persson, B.~K. Lau, E.~Larsson, T.~Marzetta, O.~Edfors, and
  F.~Tufvesson, ``Scaling up mimo: Opportunities and challenges with very large
  arrays,'' \emph{IEEE Signal Processing Magazine}, vol.~30, no.~1, pp. 40--60,
  2013.

\bibitem{Ngo13_EE_SE}
H.~Q. Ngo, E.~Larsson, and T.~Marzetta, ``Energy and spectral efficiency of
  very large multiuser mimo systems,'' \emph{Communications, IEEE Transactions
  on}, vol.~61, no.~4, pp. 1436--1449, April 2013.

\bibitem{Yang13}
H.~Yang and T.~Marzetta, ``Performance of conjugate and zero-forcing
  beamforming in large-scale antenna systems,'' \emph{Selected Areas in
  Communications, IEEE Journal on}, vol.~31, no.~2, pp. 172--179, February
  2013.

\bibitem{Fernandes13}
F.~Fernandes, A.~Ashikhmin, and T.~Marzetta, ``Inter-cell interference in
  noncooperative tdd large scale antenna systems,'' \emph{Selected Areas in
  Communications, IEEE Journal on}, vol.~31, no.~2, pp. 192--201, February
  2013.

\bibitem{Ashikhmin12}
A.~Ashikhmin and T.~Marzetta, ``Pilot contamination precoding in multi-cell
  large scale antenna systems,'' in \emph{Information Theory Proceedings
  (ISIT), 2012 IEEE International Symposium on}, July 2012, pp. 1137--1141.

\bibitem{Jose11}
J.~Jose, A.~Ashikhmin, T.~Marzetta, and S.~Vishwanath, ``Pilot contamination
  and precoding in multi-cell tdd systems,'' \emph{Wireless Communications,
  IEEE Transactions on}, vol.~10, no.~8, pp. 2640--2651, August 2011.

\bibitem{Nam13}
Y.-H. Nam, B.~L. Ng, K.~Sayana, Y.~Li, J.~Zhang, Y.~Kim, and J.~Lee,
  ``Full-dimension mimo (fd-mimo) for next generation cellular technology,''
  \emph{Communications Magazine, IEEE}, vol.~51, no.~6, pp. 172--179, June
  2013.

\bibitem{Hoydis13}
J.~Hoydis, S.~ten Brink, and M.~Debbah, ``Massive mimo in the ul/dl of cellular
  networks: How many antennas do we need?'' \emph{Selected Areas in
  Communications, IEEE Journal on}, vol.~31, no.~2, pp. 160--171, February
  2013.

\bibitem{Caire12}
H.~Huh, G.~Caire, H.~Papadopoulos, and S.~Ramprashad, ``Achieving "massive
  mimo" spectral efficiency with a not-so-large number of antennas,''
  \emph{Wireless Communications, IEEE Transactions on}, vol.~11, no.~9, pp.
  3226--3239, September 2012.

\bibitem{Ngo13}
H.~Ngo, E.~Larsson, and T.~Marzetta, ``The multicell multiuser mimo uplink with
  very large antenna arrays and a finite-dimensional channel,''
  \emph{Communications, IEEE Transactions on}, vol.~61, no.~6, pp. 2350--2361,
  June 2013.

\bibitem{Peel05}
C.~Peel, B.~Hochwald, and A.~Swindlehurst, ``A vector-perturbation technique
  for near-capacity multiantenna multiuser communication-part i: channel
  inversion and regularization,'' \emph{Communications, IEEE Transactions on},
  vol.~53, no.~1, pp. 195--202, Jan 2005.

\bibitem{Nossek05}
M.~Joham, W.~Utschick, and J.~Nossek, ``Linear transmit processing in mimo
  communications systems,'' \emph{Signal Processing, IEEE Transactions on},
  vol.~53, no.~8, pp. 2700--2712, Aug 2005.
\end{thebibliography}
\end{document}